\documentclass[aps,prl,reprint,superscriptaddress]{revtex4-1}

\usepackage{graphicx}

\usepackage{url}

\usepackage[separate-uncertainty=true,multi-part-units=single,per-mode=symbol]{siunitx}

\usepackage{amsmath}

\renewcommand{\vec}[1]{\mathbf{#1}}

\newcommand{\abs}[1]{\left\vert #1 \right\vert}
\newcommand{\order}[1]{\mathcal{O}\!\left\{ #1 \right\}}

\begin{document}

\title{Machine-learning techniques for 
  fast and accurate feature localization in
  holograms of colloidal particles}

\author{Mark D. Hannel}

\affiliation{Department of Physics 
  and Center for Soft Matter Research, 
  New York University, New York, NY 10003}

\author{Aidan Abdulali}

\affiliation{Packer Collegiate Institute, Brooklyn, NY 11201}

\author{Michael O'Brien}
\author{David G. Grier}

\begin{abstract}  
  Holograms of colloidal particles can be analyzed with the Lorenz-Mie
  theory of light scattering to measure individual particles'
  three-dimensional positions with nanometer precision
  while simultaneously estimating their sizes and refractive indexes.
  Extracting this wealth of information
  begins by detecting and localizing features of interest within
  individual holograms.  Conventionally approached with heuristic
  algorithms, this image analysis problem can be solved faster and more
  generally with machine-learning techniques.  We demonstrate that two
  popular machine-learning algorithms, cascade classifiers and deep
  convolutional neural networks (CNN), can solve the feature-localization
  problem orders of magnitude faster than current state-of-the-art
  techniques.
  Our CNN implementation localizes holographic features precisely enough
  to bootstrap more detailed analyses based on the Lorenz-Mie theory of
  light scattering.
  The wavelet-based Haar cascade proves to be less
  precise, but is so computationally
  efficient that it creates new opportunities for applications
  that emphasize speed and low cost.
  We demonstrate its use as a real-time targeting system for
  holographic optical trapping.
\end{abstract}

\maketitle



\section{Introduction: Holographic Particle Characterization}

Holographic particle characterization \cite{lee07a} uses 
quantitative analysis of holographic video microscopy images to 
measure the size, shape, and composition of individual colloidal
particles, in addition to their three-dimensional positions.
When applied to a stream of dispersed particles, holographic
characterization measurements provide insights into the
joint distribution of particle size and composition
that cannot be obtained in any other way.
This technique has been demonstrated on both homogeneous
and heterogeneous \cite{yevick14,philips17}
dispersions of colloidal spheres, and has been extended to work for 
colloidal clusters \cite{perry12,fung12,fung13}, and aggregates 
\cite{wang16,wang16a}, as well as colloidal rods \cite{cheong10} 
and other aspherical particles \cite{wang14using,hannel15}.
Applications include
monitoring protein aggregation in biopharmaceuticals \cite{wang16},
detecting agglomeration in semiconductor polishing slurries
\cite{cheong17}, 
gauging the progress of colloidal synthesis reactions \cite{wang15,wang15a},
performing microrheology \cite{cheong08}, 
microrefractometry \cite{shpaisman12}, 
and microporosimetry \cite{cheong11} measurements,
assessing the quality of dairy products \cite{cheong09a},
and monitoring contaminants 
in wastewater \cite{philips17}.

The critical first step in holographic particle characterization
is to detect features of interest within a recorded video frame,
and to localize them well enough to enable subsequent 
analysis  \cite{crocker96,cheong09,yevick14,krishnatreya14a}.
False positive and negative detections clearly are undesirable.
Poor localization slows downstream analysis
\cite{yevick14,krishnatreya14a} and can prevent fitting algorithms 
from converging to reasonable results.
Here, we demonstrate that machine-learning algorithms can meet the
need for reliable feature detection and precise object localization in
holographic video microscopy.
This complements the previously reported \cite{yevick14}
use of machine-learning regression to estimate
characteristics such as particle size from holographic
features that already have been detected, localized
and isolated by other means.
With appropriate training, machine-learning algorithms
surpass standard image-analysis
techniques in their ability to cope with common image defects
such as overlapping features.
They also operate significantly faster, thereby enabling
applications that benefit from real-time performance on low-cost
hardware.

\begin{figure}[b!]
  \centering
  \includegraphics[width=\columnwidth]{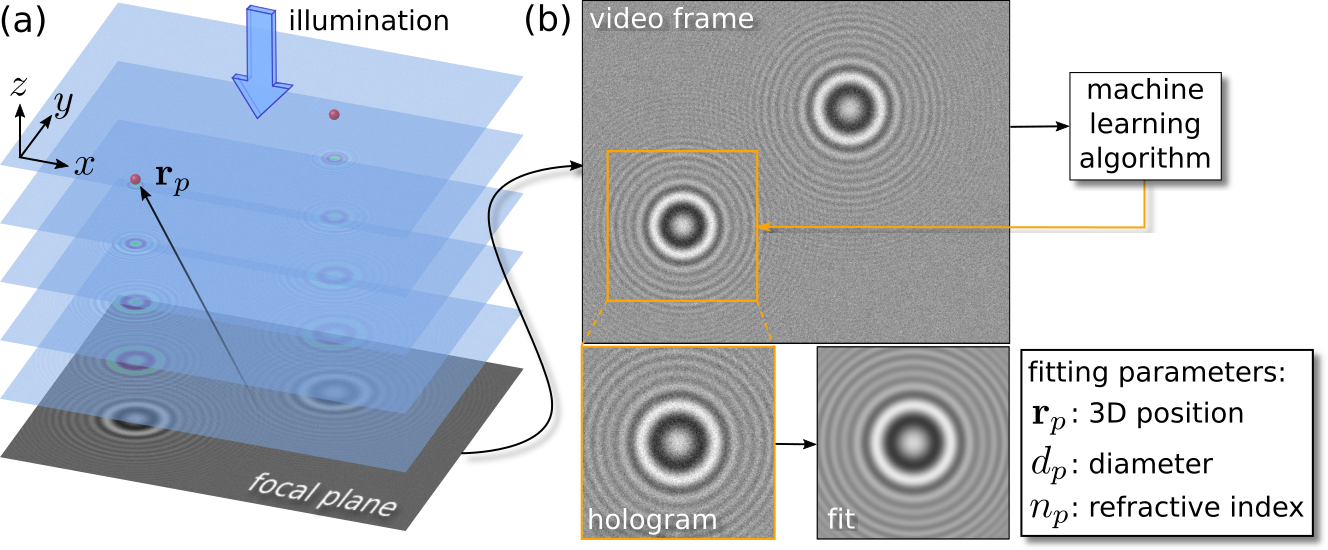}	
  \caption{Overview of holographic particle characterization. (a) Plane-wave 
  illumination is scattered by colloidal particles (red spheres). The 
  field scattered by a particle at $\vec{r}_p$ interferes with the plane wave
  to produce a hologram in the focal
  plane of a microscope.
  (b) Features in a digitally recorded hologram are
  detected with a machine-learning algorithm before being
  analyzed with light-scattering 
  theory to estimate the particles' physical properties.}
  \label{fig:characterization}
\end{figure}

\section{Detecting and Localizing Holographic Features}

Figure~\ref{fig:characterization} illustrates the challenge of
recognizing features in holograms.
Light scattered by a particle 
spreads as it propagates to the focal plane of a conventional
microscope.
There, it interferes with the remainder of the illuminating beam
to create a pattern of concentric interference fringes.
The microscope magnifies this interference pattern and
projects it onto
the detector of a video camera.
The intensity variations
associated with a single colloidal particle
typically span many pixels in a recorded image
and display rich internal structure.
Their scale and complexity render such features difficult to
recognize by conventional
particle-tracking techniques.
  
\subsection{ Heuristic Algorithms}
\label{sec:heuristic}

One practical method for detecting holographic features and 
locating their centers involves transforming extended interference
patterns into compact peaks \cite{cheong09,krishnatreya14a}, and 
then locating the peaks with standard centroid detectors 
\cite{crocker96,allan16trackpy}. 
Successful implementations of this two-step
approach have been based on
voting algorithms such as the circular Hough transform
\cite{cheong09,parthasarathy12,allan16trackpy} and
the orientation alignment transform \cite{krishnatreya14a}.
Both of these feature-coalescence algorithms rely on the
radial symmetry of typical single-particle holograms'
symmetry without reference to the underlying image-formation
mechanism.
For an image of $N$ pixels on a side, voting algorithms
have a computational complexity of $\order{N^3 \log N}$
\cite{hollitt13} whereas the convolution-based orientation alignment
transform
runs in $\order{N^2 \log N}$ operations \cite{cheong09}.
For this reason, we transform holograms with the
orientation alignment transform
to assess the performance of heuristic algorithms,
using an open-source implementation described
in Ref.~\cite{cheong09}.

The peaks created by feature coalescence can be detected
and their centroids localized as local maxima in the transformed images.
When presented with holograms of well-separated
colloidal spheres, heuristic algorithms provide
sub-pixel precision for particle localization
\cite{cheong09,krishnatreya14a}.
This easily meets the need to localize
features for subsequent analysis.
We use the open-source
TrackPy implementation \cite{allan16trackpy}
of the Crocker-Grier algorithm \cite{crocker96}.

Detecting and localizing local maxima can be very
efficient if the peaks have well-defined widths,
heights and separations \cite{crocker96,allan16trackpy}.
Transformed holograms of colloidal particles, however,
can have widely varying contrasts
and extents depending on the particles' properties and
heights above the focal plane.
Thresholds for feature detection and localization therefore
must be assessed from the transformed images
themselves.
This can create a bottleneck
for
heuristic feature detection and localization.

\subsection{Machine-Learning Algorithms}

Machine-learning techniques can reduce the computational
burden of detecting and localizing features of interest in
holographic microscopy images, and also prove to be more
robust against false positive and negative feature detections.
We have implemented two such approaches:
a cascade of boosted classifiers based on Haar-like wavelets,
and a deep convolutional neural network (CNN).
Both approaches yield estimates for the in-plane
coordinates, $(x_p, y_p)$, for every particle in the field
of view, as well as the extent of the region of interest
encompassing the scattering pattern.
The box superimposed on the two-particle hologram in
Fig.~\ref{fig:characterization}(b) represents a
region of interest centered on one of the particles
that was computed by a CNN.

Cascade classifiers and convolutional neural networks
both work by convolving holograms with
small arrays and interpreting the results.
They thus require $\order{N^2}$ operations, which
gives them the potential to run significantly more
quickly than heuristic algorithms, particularly for
larger images.
Each has particular strengths for particle localization
in holographic microscopy images.

Cascade classifiers were originally developed for detecting 
faces in photographs \cite{viola2001rapid}. 
They work by convolving an image with a sequence of 
selected wavelets, each of which is considered to be a
``weak classifier'' for objects of interest.
An above-threshold response from a linear combination of such
weak classifiers signifies the presence of a feature of interest
centered at the point of strongest response.
Regions containing such above-threshold responses are analyzed
with the weak classifiers at the next step of the cascade.
Any regions that remain after analysis by the full cascade are
considered to be features.
The analysis is performed at a sequence of resolutions to capture
features at different scales.
Haar wavelets are particularly attractive
for this application because they are implemented in
integer arithmetic with highly efficient algorithms.
The training process determines
which Haar wavelets constitute useful weak classifiers
at each level of the cascade, and which combinations best
serve as strong classifiers for features of interest.
Training also optimizes the number of stages of increasingly
fine resolution required to detect features reliably and to
localize them with a requested precision.
This approach has been adapted for a wide range of
object recognition and image segmentation tasks
\cite{lienhart2002extended}.
Our application of this technique to holographic feature 
localization is based on an open-source implementation
of Haar cascade classifiers made 
available by the OpenCV project \cite{itseez2015opencv}.
This cascade classifier can be trained to recognize non-standard
features of interest, such as holograms of colloidal particles.
For each such feature in a hologram, it yields
a candidate set of rectangular regions of interest that
may include multiple estimates for each feature.
Any such overlapping detections can be coalesced with
standard methods for non-maximum suppression
\cite{neubeck06}.
The center of each resulting rectangle constitutes an
estimate for the associated feature's position in the focal plane.

Convolutional neural networks also solve image recognition
tasks through convolutions with selected kernels. In this case,
the convolutions are integrated into the network's
multi-layered, feed-forward architecture 
\cite{sermanet2013overfeat} and employ kernels that are designed
and optimized during training.
Constructing a CNN to perform general image classification 
requires massive computational resources \cite{tensorflow2015-whitepaper}.
Once constructed, however, a CNN can be
retrained easily to recognize particular features of interest.
Our application of CNNs for feature localization is based on 
TensorBox \cite{stewart2015endtoend}, an open-source package 
built on the GoogLeNet-OverFeat network \cite{sermanet2013overfeat},
specifically on Inception v1 \cite{szegedy15}.
Tensorbox provides a convenient interface for training the
input layers of Inception to recognize features of interest
and the for training the output layers to associate these features
with regression estimates for the
locations and extents of detected features.

Both types of supervised machine-learning algorithms 
require sets of sample data for training and validation.
Each training element
consists of an image containing
zero, one or more features together with
a ``ground truth'' annotation for each feature in that image
specifying the features' locations and extents.
Normally, these images are obtained experimentally
and are annotated by hand.
We instead train with synthetic holograms that are 
computed with the same light scattering theory \cite{lee07a}
used to analyze experimental holograms.
Using the physics of image formation
for the ground truth for training eliminates 
the effort and errors inherent in empirical annotation.

\section{Holographic Image Formation}

Referring to Fig.~\ref{fig:characterization}, we model the 
holographic microscope's illumination as a
plane wave at frequency $\omega$ propagating down the $\hat{z}$ 
axis (along $-\hat{z}$) and linearly polarized along $\hat{x}$:
\begin{equation}
  \label{eq:incidentwave}
  \vec{E}_0(\vec{r},t)
  =
  u_0 \, e^{i k z} e^{i \omega t} \, \hat{x},
\end{equation}
where $k = n_m \omega/c$ is the wavenumber of the light
in a medium of refractive index $n_m$. A particle at position 
$\vec{r}_p$
scatters the incident wave, thereby creating the scattered field
\begin{equation}
  \label{eq:scatteredwave}
  \vec{E}_s(\vec{r},t)
  =
  u_0 \, e^{i k z_p} \, \vec{f}_s(k[\vec{r} - \vec{r}_p]) \, e^{i \omega t},
\end{equation}
where $\vec{f}_s(k\vec{r})$ is the Lorenz-Mie scattering 
function \cite{bohren83,mishchenko02}. For the particular case
of scattering by a sphere, $\vec{f}_s(k\vec{r})$ is parameterized
by the sphere's radius $a_p$ and refractive index $n_p$ \cite{bohren83}.
The field that reaches point $\vec{r}$ in the focal plane 
($z = 0$) is the superposition of these two contributions,
\begin{equation}
  \label{eq:modelfield}
  \vec{E}(\vec{r},t) = \vec{E}_0(\vec{r},t) + \vec{E}_s(\vec{r},t).
\end{equation}
The dimensionless intensity,
$b(\vec{r}) \equiv u_0^{-2} \abs{\vec{E}(\vec{r},t)}^2$,  
is then given by
\begin{equation}
  \label{eq:modelintensity}
  b(\vec{r})
  =
  \abs{\hat{x} + e^{i k z_p} \, \vec{f}_s(k[\vec{r} - \vec{r}_p])}^2.
\end{equation}

In addition to $a_p$ and $n_p$, this model for the
image formation process depends on a small number of
parameters that characterize the instrument.
Our holographic microscope is powered by
a \SI{15}{\milli\watt} fiber-coupled diode laser (Coherent Cube)
operating at a vacuum wavelength of 
$\lambda = \SI{447}{\nm}$.  
The combination of a half-wave plate and a polarizing 
beam splitter reduces the
power incident on the sample to \SI{3}{\milli\watt} and ensures
that the light is linearly polarized along $\hat{x}$,
as required by Eq.~\eqref{eq:modelintensity}.
A $100\times$ oil-immersion objective lens
(Nikon S-Plan Apo, numerical aperture 1.3) and a matched \SI{200}{\mm}
tube lens provide a total magnification of \SI{135}{\nm\per pixel}
on a standard video camera (NEC TI-324AII). The 
\SI{640 x 480}{pixel} grid is digitized at \SI{8}{bits \per pixel} 
and recorded as uncompressed digital video at 
\SI{29.97}{frames\per\second} with a commercial digital video 
recorder (Pioneer DVR-560H).  The refractive 
index of the medium, $n_m$, is determined to within a part 
per ten thousand using an Abbe refractometer (Edmund Scientific).

Having determined the calibration constants,
we treat the particle's position and properties as 
adjustable parameters and fit predictions of
Eq.~\eqref{eq:modelintensity} to experimentally measured holograms.
To do so, each video frame must first be 
corrected by subtracting off the camera's dark count 
\cite{wang16a}, and then
normalizing by the microscope's background intensity distribution \cite{lee07a}. 
Such fits typically yield a sphere's position with a precision of 
\SI{1}{\nm} in the plane and \SI{3}{\nm} axially 
\cite{cheong09,cheong10a}. 
Characterization results are similarly precise, 
with the radius of a micrometer-diameter sphere typically 
being resolved to within \SI{3}{\nm} and the refractive index to 
within a part per thousand \cite{krishnatreya14,shpaisman12}.

This excellent performance requires starting 
estimates for the adjustable parameters that are
good enough for the fitting 
algorithm to converge to a globally optimal solution. 
The fitter computes trial holograms according to
Eq.~\eqref{eq:modelintensity}, which is computationally 
expensive.
It has to perform fewer of these computations when it is
provided with better estimates for the starting parameters.
Whereas heurstic localization algorithms meet this need,
machine-learning algorithms are substantially faster and
more robust, and can be comparably precise.

\section{Applying Machine Learning 
to Holographic Particle Localization}

We used Eq.~\eqref{eq:modelintensity} to generate training 
images of particles with radii ranging from 
$a_p = \SI{0.25}{\um}$ to \SI{5}{\um}, refractive indexes from 
$n_p = \num{1.4}$ to \num{2.5}, and axial positions from 
$z_p = \SI{5}{\um}$ to \SI{50}{\um}. 
Each training hologram has parameters selected at 
random from this 
range and is centered at random within
the field of view.
Normalized experimental holograms have uncorrelated white
noise that we model as additive Gaussian noise with a standard
deviation of five percent.

Our cascade classifier was trained with \num{6000}
synthetic images of colloidal spheres.
These were combined with a complementary set of 
\num{4000} particle-free images recorded by the instrument itself.
Each computed image is annotated with
the coordinates of the corners defining that feature's region of
interest.
The region is centered on the feature's actual position
and has an extent that encloses 10 interference fringes.
The classifier was trained
until its rate of false positive detections
fell to \num{8E-4}.
This was achieved with a classifier that searches for features
through five resolution stages, with each stage being
comprised of a distinct set of five wavelets.
This geometry and the choice of weak classifiers was arrived at
by the training algorithm's optimizer.

The convolutional neural network was trained with 
\num{3000} synthetic holographic images; another \num{600} 
were used for validation.
These images also were annotated with feature positions and extents
drawn from the ground truth for the image-formation process.
CNN training converged after \num{50000} cycles
of training and validation.

\subsection{Precision and Accuracy}
  
\begin{figure}[!t]
  \centering
  \includegraphics[width=\columnwidth]{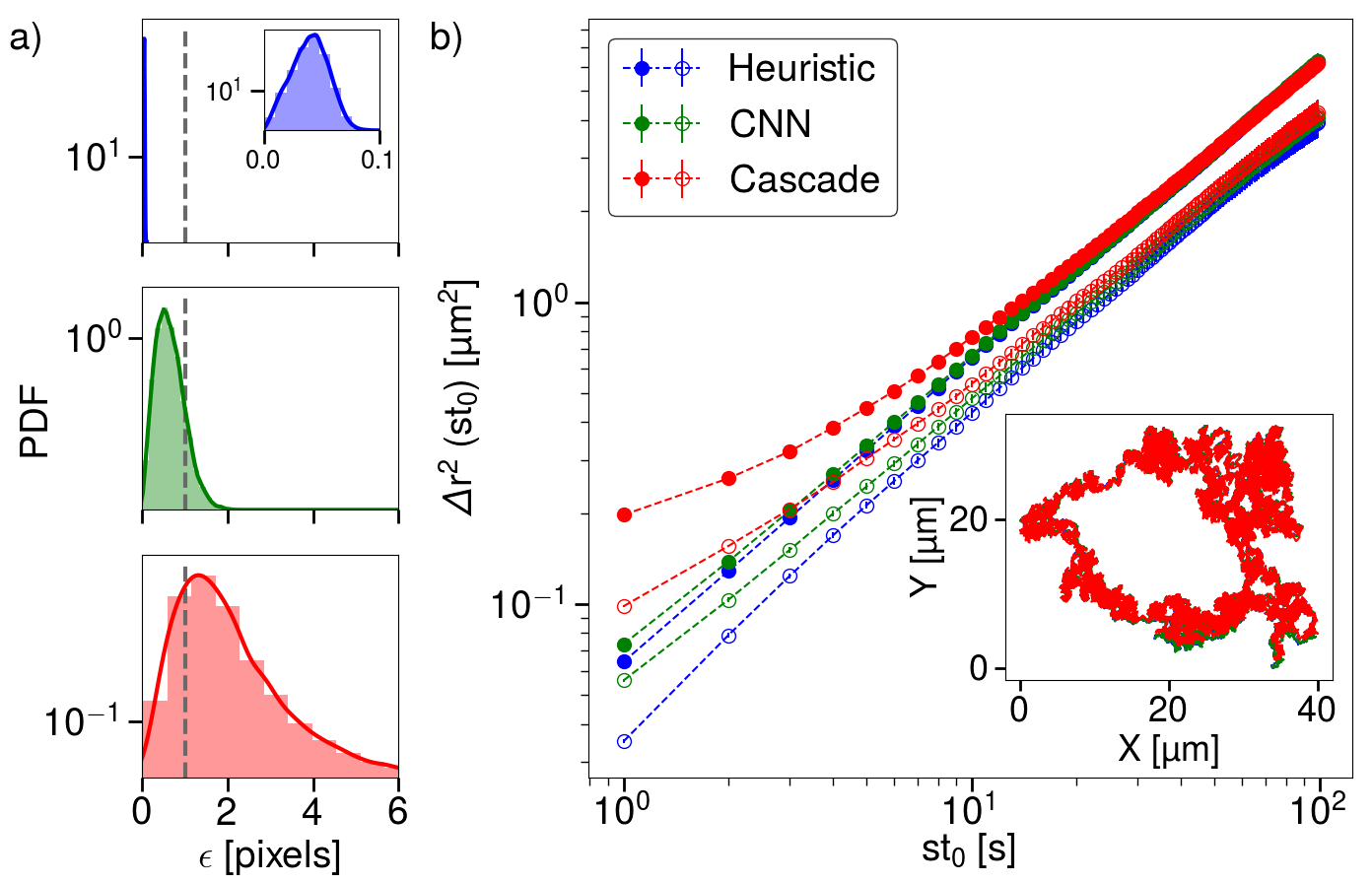}
  \caption{Each localization technique provided estimates
  for the trajectory of a simulated brownian particle.
  (a) Probability distribution functions for the
  localization error achieved by (top) heuristic
  algorithm, (middle) convolutional neural network,
  and (bottom) cascade classifier. 
  Inset shows expanded view of the subpixel
  resolution. Vertical dashed line indicates single-pixel
  precision.
  (b) Mean-square displacement computed from 
  trajectories obtained with the three detection algorithms.
  Short-time asymptotes yield dynamical estimates
  for the localization error.
  Open circles represent experimental data,
  as explained in Sec.~\ref{sec:experiment}.
  }
  \label{fig:msdplot}
\end{figure}

\begin{figure}[!t]
  \centering
  \includegraphics[width=\columnwidth]{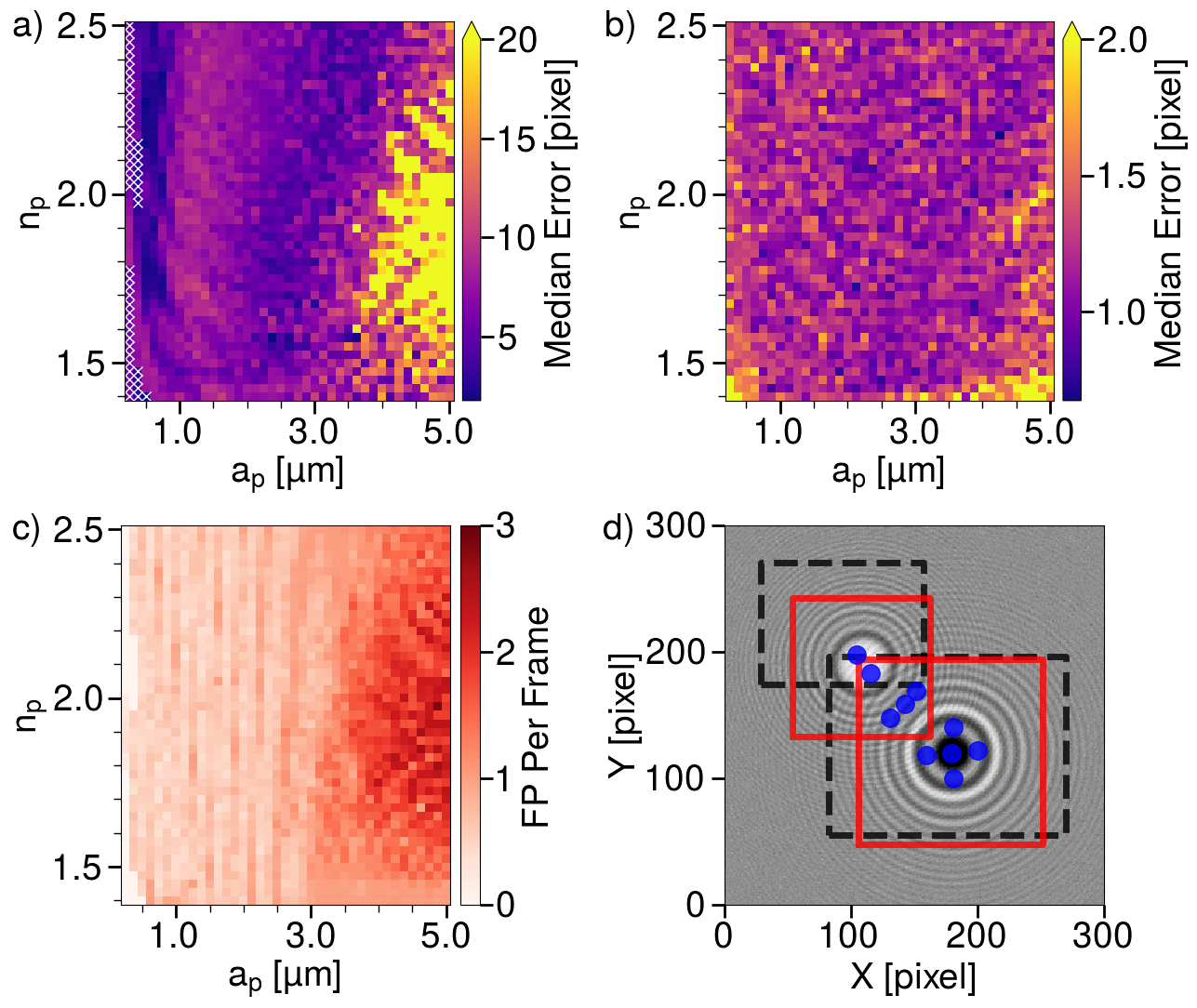}
  \caption{Localization errors as a function of particle
  radius and refractive index at a height of $z_p = \SI{13.5}{\um}$
  above the focal plane. (a) Cascade classifier. (b) Convolutional
  neural network. (c) Rate of false positive detections for the 
  cascade classifier. (d) The hologram of a \SI{2.4}{\um}-diameter
  polystyrene sphere (upper left) interfering with the hologram of
  a \SI{4.0}{\um}-diameter TPM sphere located \SI{15}{\um} above it.
  Blue dots show feature locations proposed by the orientation 
  alignment transform; Red boxes enclose features detected by the CNN; 
  Dashed black boxes are proposed by the cascade classifier.
  }
  \label{fig:figure3}
\end{figure}

We assess the detectors' localization precision by 
comparing detection results with known 
input parameters.
A typical example for a particular choice of particle
properties is shown in Fig.~\ref{fig:msdplot}.
The three probability distributions 
in Fig.~\ref{fig:msdplot}(a) present
the root-mean-square localization error
obtained by each of the algorithms when tracking
particles with
$a_p =\SI{1.0}{\um}$ and $n_p = \num{1.5}$.
We generate data for these plots by simulating the
diffusion of such a particle through water
at a temperature of \SI{20}{\celsius} starting from 
the center of the field of view at $z_p=\SI{13.5}{\um}$
and proceeding for \num{3000} steps at \SI{33}{\ms} per step.

The heuristic algorithm consistently 
yields sub-pixel precision with a median error of 
\SI{0.04}{pixels}. 
The convolutional neural network also yields sub-pixel
precision with a median localization error of \SI{0.61}{pixels}.
The cascade classifier performs less well,
with a median localization error of \SI{1.81}{pixels} and
a substantial probability
for errors extending to several pixels.
For applications such as Lorenz-Mie microscopy
that require input estimates with sub-pixel precision,
the cascade classifier's localization precision may not
be sufficient.

The inset of Fig.~\ref{fig:msdplot}(b) shows the trajectory
reconstructed by each of the algorithms.
The measured trajectory's mean-squared displacement (MSD)
provides an estimate for the particle's
diffusion coefficient.
All three methods yield results that are consistent with
the particle's true diffusivity,
$D = \SI{0.482}{\um^2 \per \second}$,
which suggests that
their localization errors are normally distributed.
Extrapolating the MSD to zero lag time 
provides an estimate for the localization error \cite{crocker96,michalet12}.
In all three cases, the extrapolated measurement error
is consistent with the median values from Fig.~\ref{fig:msdplot}(a). 

Applying the same techniques across the entire range of particle
sizes and refractive indexes yields results for the median
localization error summarized in
Fig.~\ref{fig:figure3}(a) and \ref{fig:figure3}(b).
Results from the cascade classifer in Fig.~\ref{fig:figure3}(a)
range from single-pixel precision under most conditions to 
more than 20 pixels for the largest spheres we considered.
These errors are dominated by the cascade classifier's tendency
to displace location estimates toward the center of the
field of view when presented with features that extend outside
the observation window.
This problem is more pronounced
for the larger holographic features created by larger scatterers.
Smaller particles create holograms with low signal-to-noise 
ratio that can be overlooked by the cascade classifier, leading
to false negative detections.
Such conditions are indicated by white crosses in
Fig.~\ref{fig:figure3}(a).

The results plotted in Fig.~\ref{fig:figure3}(b) show that
the CNN yields much smaller
localization errors than the cascade classifier.
The CNN achieves sub-pixel
resolution over the entire range of parameters,
although localization precision is worse for
weak scatterers and large spheres.
Unlike the cascade classifier, it also returned no
false negative results.

Both the cascade classifier and the CNN
return a small rate of false positive detections.
Figure~\ref{fig:figure3}(c) reports the false-positive
rate for the cascade classifier, which ranges from
\SI{E-1}{frame^{-1}} for holograms of particles with
$a_p < \SI{3}{\um}$
to \SI{3}{frame^{-1}} for holograms of larger spheres.
In all cases, these false positive detections come in addition
to the correct particle detection, and result from
the classifier's failure to correct coalesce multiple detections
of the same particle.
Such false positive detections contribute to the very large
localization error for large spheres in Fig.~\ref{fig:figure3}(a).
The CNN performs substantially
better, with fewer than one false positive per thousand
frames.

\subsection{Multiple Particles}

The results presented so far apply to holograms of single
particles.
In practice, it is not unusual for multiple particles to enter 
the microscope's field of view simultaneously.
Their scattering patterns interfere to create
intensity variations that can confuse heuristic detection algorithms.
Depending on the particles' proximity and alignment, their
holograms can merge into irregular patterns whose analysis
requires more specialized techniques \cite{perry12,fung13}.
The hologram in Fig.~\ref{fig:figure3}(d)
illustrates the effect of more modest overlap.
It captures a \SI{2.4}{\um}-diameter
polystyrene sphere \SI{17}{\um} above the focal plane
whose hologram is partially occluded by that 
of a \SI{4.0}{\um}-diameter TPM sphere situated \SI{15}{\um}
above and \SI{15}{\um} off to the side.
Discrete points overlaid on this image show the
positions that the heuristic algorithm
identified as centers of candidate features.
Of the 10 proposed features, 8 are false positive detections
and one is poorly localized.

Both machine-learning algorithms perform better than the
heuristic algorithm for this image.
The cascade classifier correctly
detects both particles, as indicated by dashed rectangles
in Fig.~\ref{fig:figure3}(d).
The estimated locations, however, are displaced significantly
from the features' true centers, presumably because of
interference between the two scattering patterns.
The CNN not only detects and localizes
both particles correctly, but also provides useful estimates for the
extent of the scattering patterns, as denoted by the solid (red)
squares overlaid on Fig.~\ref{fig:figure3}(d).

These results illustrate that machine-learning algorithms can
be more reliable than heuristic algorithms for detecting and
localizing features in non-ideal holograms.
For applications such as monitoring colloidal concentrations, this
benefit alone might recommend machine-learning algorithms
over other approaches.
The principal benefit of machine-learning algorithms, however,
is their ability to detect features rapidly, even on
low-power computational platforms.

\subsection{Computation Speed}

Table~\ref{table:times} presents timing data for holographic
feature detection on a \SI{1}{Gflops} desktop workstation
outfitted with an
nVidia GTX 680 GPU.
This system can detect a single feature in just under
\SI{700}{\ms} using the heuristic
algorithm described in Sec.~\ref{sec:heuristic}.
Of this, \SI{150}{\ms} is required for the orientation alignment
transform and half a second is required to analyze the
transformed image and then to detect and localize
its peaks.
This bottleneck can be reduced to \SI{50}{\ms} by specifying
the anticipated width, height and separation of the
transformed peaks.
In this case, the present implementation's processing speed
is consistent with previous reports \cite{lee07a,cheong09,allan16trackpy}
when account is taken of image size and processor speed.
No single set of such parameters, however, successfully detects
features over the entire range of parameters considered in
Fig.~\ref{fig:figure3}.
The slow operation reported in Table~\ref{table:times}
therefore represents the cost of generality.

The CNN routinely
outperforms the heuristic algorithm by
a factor of \num{2.5} on the same hardware over the entire
range of parameters.
Transferring the CNN calculation to the GPU
increases this advantage to a factor of \num{11}.
Most remarkably, the cascade classifier is \num{40} times
faster than the reference heuristic algorithms,
even without GPU acceleration, processing features
fast enough to keep up with the \SI{33}{\ms} frame
rate of a standard video camera.

\begin{table*}
\centering
\caption{Analysis times in ms/frame for the heuristic
  algorithm, the convolutional neural network (CNN) implemented on
  CPU and GPU, and the cascade classifier implemented on a
  workstation and on a Raspberry Pi 3 single-board computer.}
\begin{tabular}{lrrrrr}
\hline
\hline
  & \text{Mean~[ms]} & \text{Median~[ms]} & \text{Std.~[ms]} &
                                                               \text{Min~[ms]} & \text{Max~[ms]} \\
Heuristic (CPU) & 695 & 700 & 11 & 670 & 1000 \\ 
CNN (CPU) & 278 & 278 & 2.8 & 271 & 315 \\
CNN (GPU) & 52 & 52 & 4.8 & 50 & 70 \\
Cascade (CPU) & 17 & 17 & 1.0 & 15 & 81 \\ 
Cascade (RPi) & 173 & 171 & 12 & 159 & 275 \\ \hline \hline
\end{tabular}
\label{table:times}
\end{table*}

The cascade classifier is so computationally efficient
that it can be deployed usefully on a lightweight
embedded computer. 
We demonstrated this by analyzing holograms on
a Raspberry Pi 3 single-board computer.
Even though the light-weight computer runs the
cascade classifier \num{10} times slower than the
workstation,  it is still \num{4} times faster than the heuristic 
algorithms on the workstation.
Reducing the resolution by half, improves the 
Raspberry Pi's detection time to \SI{40}{\ms} per image 
which corresponds to \SI{25}{frames \per \second}.

\subsection{Experimental Demonstrations}
\label{sec:experiment}

\begin{figure}
  \centering
  \includegraphics[width=\columnwidth]{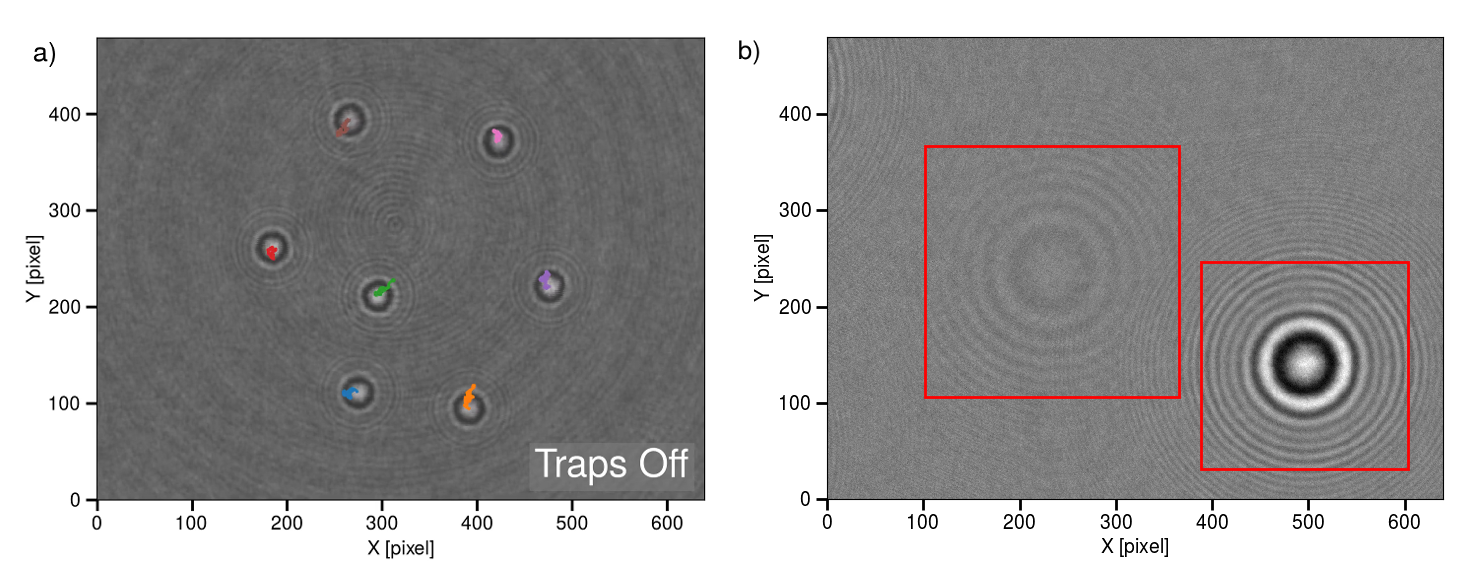}
  \caption{(a) Cascade classifier tracking \SI{2}{\um}-diameter
    colloidal spheres diffusing through water in a
    holographic optical trapping system.
    Each trace shows 5 seconds of the associated particle's motion.
    (b) CNN detection of holographic features.
    The high-contrast feature is created by a \SI{1.5}{\um}
    diameter silica sphere.  The low-contrast feature
    represents a coliform bacterium in the dispersion.}
  \label{fig:autotrap}
\end{figure}

Real-time detection of holographic features has applications
beyond holographic particle characterization.
The implementations presented here are suitable for targeting
optical traps in holographic trapping systems \cite{grier03}.
We have demonstrated this by integrating machine-learning
particle detection into an automated trapping system that
projects optical traps onto the particles' positions to acquire
them for subsequent processing.
Pioneering implementations
of automated trapping \cite{chapin06} rely on conventional
imaging and so require target particles to lie near the microscope's
focal plane.
Holographic targeting works over a much larger axial
range.
Both the CNN and the cascade classifier
locate particles in the plane with sufficient precision to ensure
reliable trapping.
The axial coordinate required for three-dimensional targeting can
be extracted from holographic features using previously reported
techniques \cite{yevick14}.
Because of its speed, the cascade classifier is particularly useful
for targeting fast-moving particles.
Figure~\ref{fig:autotrap}(a) shows the cascade classifier tracking colloidal
particles in real time as they diffuse in a holographic trapping
system.
The instrument uses this tracking information to trap the
detected particles, as shown in the associated video
(Visualization 1).

Figure~\ref{fig:autotrap}(b) shows typical results obtained with
the CNN analyzing experimental data.  The image
is a normalized hologram of a
\SI{1.5}{\um}-diameter silica sphere dispersed in water
flowing down a \SI{100}{\um}-deep microfluidic channel.
The second holographic feature in this image is due
to a coliform bacterium in the sample \cite{philips17}.
The CNN detects and correctly localizes
both features despite their substantial
difference in contrast.

We can estimate the feature-detection algorithms' precision
for particle localization by tracking diffusing particles
\cite{crocker96}.
The open circles in Fig.~\ref{fig:msdplot}(b) show results obtained
with heuristic and machine-learning algorithms for the
mean-square displacement of a single colloidal polystyrene sphere
(Duke Scientific, catalog no.\ 4016A, nominal diameter
\SI{1.587+-0.018}{\um}) diffusing through water at
room temperature.
Because polystyrene is \SI{5}{\percent} more dense than water,
the particle sediments \SI{11}{\um} over the course of this
\SI{3}{\minute} measurement.
In all three cases, the in-plane localization error obtained
by extrapolating these results is consistent with that
reported for synthetic data.

\section{Discussion}

The use of machine-learning algorithms for detecting and
localizing holographic features enable and enhance a host of 
applications for holographic video microscopy.
CNNs detect and localize colloidal
particles faster than conventional image-analysis techniques
and localize particles well enough for subsequent processing.
Our implementation also estimates the extent of each holographic feature 
thereby bypassing the standard next step in Lorenz-Mie microscopy
\cite{cheong09} and saving additional time.
These substantial speed enhancements make it possible to
perform holographic particle characterization measurements in real
time rather than requiring off-line processing.
CNNs also are more successful at
interpreting overlapping features in multi-particle
holograms and thus can be used to analyze more concentrated
suspensions.

The Haar-based cascade classifier also outstrips the heuristic algorithms'
ability to detect colloidal particles, particularly in heterogeneous
samples and crowded fields of view.
Although it cannot match the localization precision of CNNs,
its speed and modest computational requirements
create new opportunities.
We have deployed our cascade classifier on a light-weight
single-board computer and have demonstrated its utility for counting particles
and thus for measuring colloidal concentrations.
Such a low-cost instrument should be useful for routine monitoring of
industrial processes and products and for environmental monitoring.
We also have demonstrated the cascade classifier's utility for
high-speed targeting in holographic trapping.  In this case, speed
is more important than localization precision for interacting with
processes as they unfold.

While the present study focuses on detecting and localizing
holographic features with radial symmetry,
the machine-learning framework can be applied equally well to
asymmetric holograms produced by rods, clusters
or biological samples.
By reducing the computational burden of analyzing holograms,
machine-learning algorithms extend the reach of holographic
tracking and holographic characterization.
More generally, machine-learning algorithms are well-suited to bootstrapping
the more detailed analysis involved in holographic particle
characterization.
We anticipate that more of these physics-based
processing steps will be taken over by machine-learning algorithms
as that technology advances.

Open-source software for holographic particle tracking and characterization
is available online at \url{https://github.com/davidgrier/}.

\section*{Funding}

This work was supported primarily by the MRSEC program of the
National Science Foundation through Award no.\ DMR-1420073
and in part by the SBIR program of the NSF through Award no.\
IPP-1519057.
The holographic trapping instrument was developed under the
MRI program of the NSF through Grant Number DMR-0922680.

\end{document}